\documentclass[vecphys]{svmult}

\usepackage{makeidx}         
\usepackage{graphicx}        

\usepackage{multicol}        
\usepackage[bottom]{footmisc}

\makeindex             

\begin{document}

\title*{Microquasars: summary and outlook}
\author{I.F. Mirabel}
\institute{Laboratoire AIM,Irfu/Service d'Astrophysique, Bat. 709, CEA-Saclay, 91191 Gif-sur-Yvette Cedex, France and Instituto de Astronom\'\i a y F\'\i sica del Espacio (IAFE), CC 67, Suc. 28, 1428 Buenos Aires, Argentina  \\
\texttt{felix.mirabel@cea.fr}}

\maketitle

\abstract{
Microquasars are compact objects (stellar-mass black holes and neutron stars) that mimic, on a smaller scale, many of the phenomena seen in quasars. Their discovery provided new insights into the physics of relativistic jets observed elsewhere in the universe, and in particular, the accretion--jet coupling in black holes. Microquasars are opening new horizons for the understanding of ultraluminous X-ray sources observed in external galaxies, gamma-ray bursts of long duration, and the origin of stellar black holes and neutron stars. Microquasars are one of the best laboratories to probe General Relativity in the limit of the strongest gravitational fields, and as such, have become an area of topical interest for both high energy physics and astrophysics. At present, back hole astrophysics exhibits historical and epistemological similarities with the origins of stellar astrophysics in the last century.
}

\section{Introduction}
\label{sec:1}

\begin{figure}
\centering
\includegraphics[height=9cm]{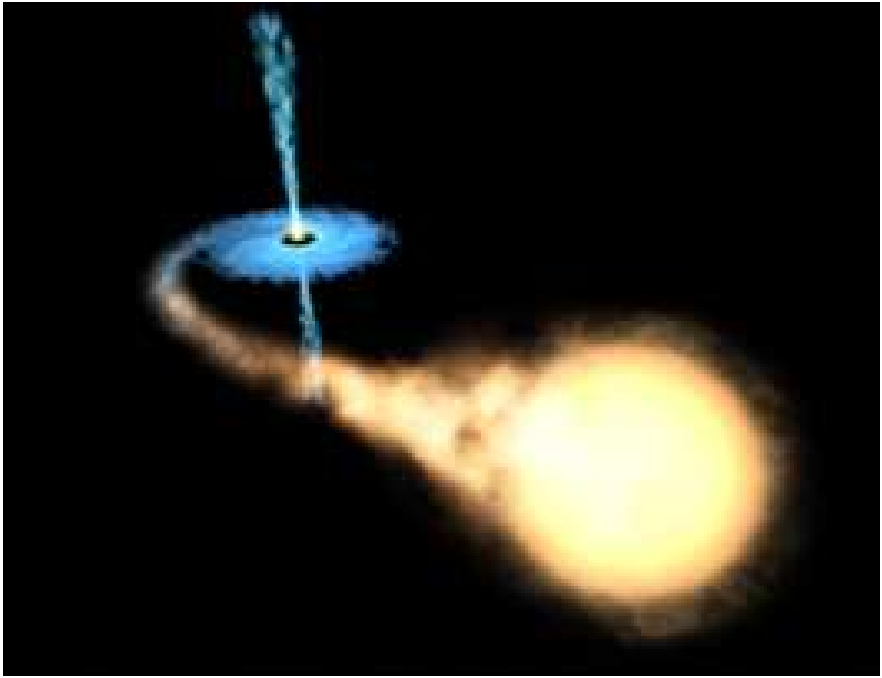}
\caption{
In our galaxy there exist binary stellar systems where an ordinary star gravitates around a black hole that sucks the outer layers of the star's atmosphere. When falling out to the dense star, the matter warms and emits huge amounts of energy as X- and $\gamma$-rays. The accretion disk that emits this radiation also produces relativistic plasma jets all along the axis of rotation of the black hole. The physical mechanisms of accretion and ejection of matter are similar to those found in quasars, but in million times smaller scales. Those miniature versions of quasars are known under the name of `microquasars'. 
}
\label{fig:1}
\end{figure}

Microquasars are binary stellar systems where the remnant of a star that has collapsed to form a dark and compact object (such as a neutron star or a black hole) is gravitationally linked to a star that still produces light, and around which it makes a closed orbital movement. In this cosmic dance of a dead star with a living one, the  first sucks matter from the second, producing radiation and very high energetic particles (Fig. \ref{fig:1}). These binary star systems in our galaxy are known under the name of ``microquasars'' because they are miniature versions of the quasars (`quasi-stellar-radio-source'), that are the nuclei of distant galaxies harboring a super massive black hole, and are able to produce in a region as compact as the solar system, the luminosity of 100 galaxies like the Milky Way. Nowadays the study of microquasars is one of the main scientific motivations of the space observatories that probe the X-ray and $\gamma$-ray Universe. 

Despite of the differences in the involved masses and in the time and length scales, the physical processes in microquasars are similar to those found in quasars. That is why the study of microquasars in our galaxy has enabled a better understanding of what happens in the distant quasars and AGN. Moreover, the study of microquasars may provide clues for the understanding of the class of gamma-ray bursts that are associated to the collapse of massive stars leading to the formation of stellar black holes, which are the most energetic phenomena in the Universe after the Big-Bang. 

\section{Discovery of microquasars}
\label{sec:2}

During the second half of the 18th century, John Michell and Pierre-Simon Laplace first imagined compact and dark objects in the context of the classical concept of gravitation. In the 20th century in the context of Einstein's General Relativity theory of gravitation, those compact and dark objects were named black holes. They were then identified in the sky in the 1960s as X-ray binaries. Indeed, those compact objects, when associated to other stars, are activated by the accretion of very hot gas that emits X and $\gamma$-rays. In 2002, Riccardo Giacconi was awarded the Nobel Prize for the development of the X-ray Space Astronomy that led to the discovery of the first X-ray binaries \cite{Giacconi}.  Later, Margon et al. \cite{Margon} found that a compact binary known as SS~433\index{SS~433} was able to produce jets of matter. However, for a long time, people believed that SS~433 was a very rare object of the Milky Way and its relation with quasars was not clear since the jets of this object move only at 26\% of the speed of light, whereas the jets of quasars can  move at speeds close to the speed of light. 

In the 1990s, after the launch of the Franco--Soviet satellite GRANAT\index{GRANAT}, growing evidences of the relation between relativistic jets and X-ray binaries began to appear. The on-board telescope SIGMA\index{SIGMA} was able to take X-ray and $\gamma$-ray images. It detected numerous black holes in the Milky Way. Moreover, thanks to the coded-mask-optics, it became possible for the first time to determine the position of $\gamma$-ray sources with arcmin precision. This is not a very high precision for astronomers who are used to dealing with other observing techniques. However, in high-energy astrophysics it represented a gain of at least one order of magnitude. It consequently made possible the systematic identification of compact $\gamma$-ray sources at radio, infrared and visible wavelengths. 

With SIGMA/GRANAT it was possible to localize with an unprecedented precision the hard X-ray and $\gamma$-ray sources. In order to determine the nature of those X-ray binaries, a precision of a few tens of arc-seconds was needed. Sources that produce high energy photons should also produce high energy particles, that should then produce synchrotron radiation when accelerated in magnetic fields. Then, with Luis Felipe Rodr\'\i guez, we performed a systematic search of synchrotron emissions from X-ray binaries with the Very Large Array (VLA) of the National Radio Astronomy Observatory of the USA.

\begin{figure}
\centering
\includegraphics[height=9cm]{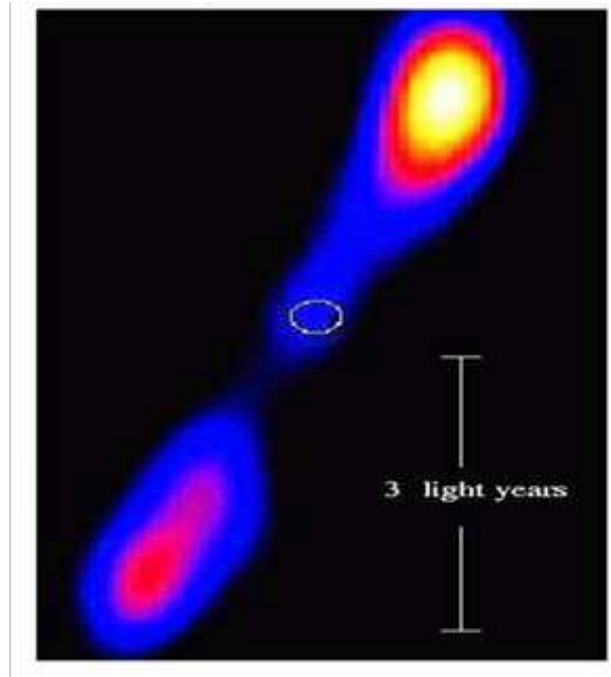}
\caption{
The British journal Nature announced the 16th of July, 1992 the discovery of a microquasar in the galactic centre region \cite{Mirabel1740}. The image shows the synchrotron emission at a radio wavelength of 6cm produced by relativistic particles jets ejected from some tens of kilometers to light-years of distance from the black hole binary which is located inside the small white ellipse.  
}
\label{fig:2}
\end{figure}

In 1992, using quasi-simultaneous observations from space with GRANAT and from the ground with the VLA, we determined the position of the radio counterpart of an X-ray source named 1E~1740.7-2942\index{1E~1740.7-2942} with a precision of sub-arc-seconds. With GRANAT this object was identified as the most luminous, persistent source of soft $\gamma$-rays in the Galactic center region. Moreover, its luminosity, variability and the X-ray spectrum were consistent with those of an accretion disk gravitating around a stellar mass black hole, like in Cygnus~X-1\index{Cyg~X-1}. The most surprising finding with the VLA was the existence of well collimated two-sided jets that seem to arise from the compact radio counterpart of the X-ray source \cite{Mirabel1740}. These jets of magnetized plasma had the same morphology as the jets observed in quasars and radio galaxies. When we published those results, we employed the term microquasar to define this new X-ray source with relativistic jets in our Galaxy. This term appeared on the front page of the British journal Nature (see Fig. \ref{fig:2}), which provoked multiple debates. Today the concept of microquasar is universally accepted and used widely in scientific publications.

Before the discovery of its radio counterpart, 1E~1740.7-2942\index{1E~1740.7-2942} was suspected to be a prominent source of 511 keV electron-positron annihilation radiation observed from the centre of our Galaxy \cite{Leventhal}, and for that reason it was nicknamed as the ``Great Annihilator''.  It is interesting that recently it was  reported \cite{Weiden} that the  distribution in the Galactic disk of the 511 keV emission, due to positron-electron annihilation, exhibit similar asymmetric distribution as that of the hard low mass X-ray binaries, where the compact objects are  believed to be stellar black holes.  This finding suggests that black hole binaries may be  important sources of positrons that would annihilate with electrons in the interstellar medium.  Therefore, positron-electron pairs may be produced by $\gamma$--$\gamma$ photon interactions in the inner accretion disks, and microquasar jets would contain positrons as well as electrons. If this recent  report is confirmed, 1E~1740.7-2942\index{1E~1740.7-2942} would be the most prominent compact source of anti-matter in the Galactic Centre region.

\section{Discovery of superluminal motions}
\label{sec:3}

If the proposed analogy \cite{MirabelRodriguez1998} between microquasars and quasars was correct, it should be possible to observe superluminal apparent motions in Galactic sources. However, superluminal apparent motions had been observed only in the neighborhood of super-massive black holes in quasars. In 1E~1740.7-2942\index{1E~1740.7-2942} we could not be able to discern motions, as in that persistent source of $\gamma$-rays the flow of particles is semi-continuous. The only possibility of knowing if superluminal apparent movements exist in microquasars was through the observation of a discreet and very intense ejection in an X-ray binary. This would allow us to follow the displacement in the firmament of discrete plasma clouds. Indeed, with the GRANAT\index{GRANAT} satellite was discovered \cite{Castro1915} a new source of X-rays with such characteristics denominated GRS~1915+105\index{GRS~1915+105}. Then with Rodr\'\i guez we began with the VLA a systematic campaign of observations of that new object in the radio domain, and in collaboration with Pierre-Alain Duc (CNRS-France) and Sylvain Chaty (Paris University) we performed the follow-up of this source in the infrared with telescopes of the Southern European Observatory, and telescopes at Mauna Kea, Hawaii.

Since the beginning, GRS~1915+105\index{GRS~1915+105} exhibited unusual properties. The observations in the optical and the infrared showed that this X-ray binary was very absorbed by the interstellar dust along the line of sight in the Milky Way, and that the infrared counterpart was varying rapidly as a function of time. Moreover, the radio counterpart seemed to change its position in the sky, so that at the beginning we did not know if those changes were due to radiation reflection or refraction in an inhomogeneous circumstellar medium (``Christmas tree effect''), or rather due to the movement at very high speeds of jets of matter. For two years we kept on watching this X-ray binary without exactly understanding its behavior. However, in March 1994, GRS~1915+105\index{GRS~1915+105} produced a violent eruption of X and $\gamma$-rays, followed by a bipolar ejection of unusually bright plasma clouds, whose displacement in the sky could be followed during 2 months. From the amount of atomic hydrogen absorbed in the strong continuum radiation we could infer that the X-ray binary stands at about 30000 light years from the Earth. This enabled us to know that the movement in the sky of the ejected clouds implies apparent speeds higher than the speed of light. 

\begin{figure}
\centering
\includegraphics[height=9cm]{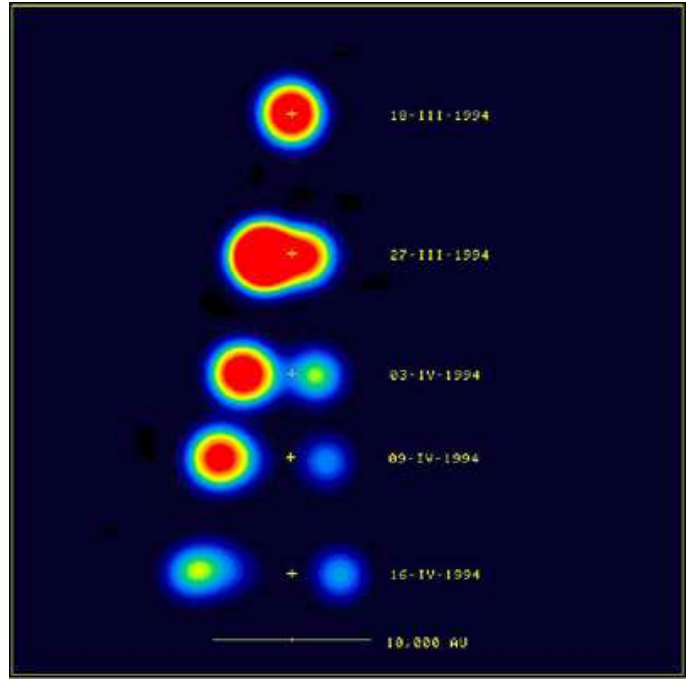}
\caption{
The journal Nature announces the 1st of September, 1994 the discovery of the first Galactic source of superluminal apparent motions \cite{MirabelRodriguez1994}. The sequence of images shows the temporal evolution in radio waves at a wavelength of 3.6 cm of a pair of plasma clouds ejected from black hole surroundings at a velocity of 98\% the speed of the light. 
}
\label{fig:3}
\end{figure}

The discovery of these superluminal apparent movements in the Milky Way was announced in Nature \cite{MirabelRodriguez1994} (Fig. \ref{fig:3}). This constituted a full confirmation of the hypothesis, that we had proposed two years before, on the analogy between microquasars and quasars. With Rodr\'\i guez we formulated and solved the system of equations that describe the observed phenomenon. The apparent asymmetries in the brightness and the displacement of the two plasma clouds could naturally be explained in terms of the relativistic aberration in the radiation of twin plasma clouds ejected in an antisymmetric way at 98\% of the speed of light \cite{MirabelRodriguez1999}. The super-luminal motions observed in 1994 with the VLA \cite{MirabelRodriguez1994} were a few years later re-observed  with higher angular resolution using the MERLIN array \cite{Fender1999}.

Using the Very Large Telescope of the European Southern Observatory, it was possible to determine the orbital parameters of GRS~1915+105, concluding that it is a binary system constituted by a black hole of $\sim$14 solar masses accompanied by a star of 1 solar mass \cite{Greiner}. The latter has become a red giant from which the black hole sucks matter under the form of an accretion disk (see Fig. \ref{fig:1}). 

\section{Disk--jet coupling in microquasars}
\label{sec:4}

The association of bipolar jets and accretion disks seems to be a universal phenomenon in quasars and microquasars. The predominant idea is that matter jets are driven by the enormous rotation energy of the compact object and accretion disk that surrounds it. Through magneto-hydrodynamic mechanisms, the rotation energy is evacuated through the poles by means of jets, as the rest can fall towards the gravitational attraction centre. In spite of the apparent universality of this relationship between accretion disks and bipolar, highly collimated jets, the temporal sequence of the phenomena had never been observed  in real time. 

Since the scales of time of the phenomena around black holes are proportional to their mass, the accretion-ejection coupling in stellar-mass black holes can be observed in intervals of time that are  millions of time smaller than in AGN and quasars. Because of the proximity, the frequency and the rapid variability of energetic eruptions, GRS~1915+105\index{GRS~1915+105} became the most adequate object to study the connection between instabilities in the accretion disks and the genesis of bipolar jets.  

\begin{figure}
\centering
\includegraphics[height=9cm]{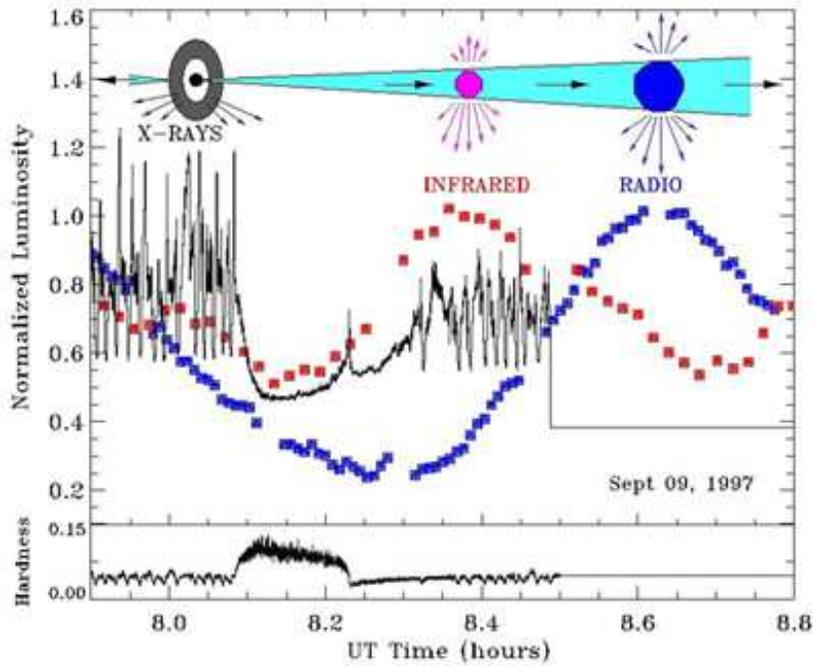}
\caption{
Temporal sequence of accretion disk -- jet coupling observed for first time in real time simultaneously in the X-rays, the infrared and radio wavelengths in the microquasar GRS~1915+105\index{GRS~1915+105} \cite{Mirabel1998}. The ejection of relativistic jets takes place after the evacuation and/or dissipation of matter and energy, at the time of the reconstruction of the inner side of the accretion disk, corona or base of the jet. A similar process has been observed years later in quasars \cite{Marscher}, but on time scales of years. As expected in the context of the analogy between quasars and microquasars \cite{MirabelRodriguez1998}, the time scale of physical processes in the surroundings of black holes is proportional to their masses.  
}
\label{fig:4}
\end{figure}

After several attempts, finally in 1997 we could observe \cite{Mirabel1998} on an interval of time shorter than an hour, a sudden fall in the luminosity in X and soft $\gamma$-rays, followed by the ejection of jets, first observed in the infrared, then at radio frequencies  (see Fig. \ref{fig:4}). The abrupt fall in X-ray luminosity could be interpreted as the silent disappearance of the warmer inner part of the accretion disk beyond the horizon of the black hole. A few minutes later, fresh matter coming from the companion star come to feed again the accretion disk, which must evacuate part of its kinetic energy under the form of bipolar jets. When moving away, the plasma clouds expand adiabatically, becoming more transparent to its own radiation, first in the infrared and then in radio frequency. The observed interval of time between the infrared and radio peaks is consistent with that  predicted  by van der Laan \cite{vdlaan} for extragalactic radio sources. 

Based on the observations of GRS~1915+105\index{GRS~1915+105} and other X-ray binaries, it was proposed \cite{fbg}  proposed a unified semiquantitative model for disk-jet coupling in black hole X-ray binary systems that relate different X-ray sates with radio states, including the compact, steady jets associated to low-hard X-ray states, that had been imaged \cite{Dhawan} using the Very Long Baseline Array of the National Radio Astronomy Observatory.

After three years of multi-wavelength monitoring an analogous sequence of X-ray emission dips followed by the ejection of bright super-luminal knots in radio jets was reported \cite{Marscher} in the active galactic nucleus of the galaxy 3C~120\index{3C~120}. The mean time between X-ray dips was of the order of years, as expected from scaling with the mass of the black hole. 

\section{Can we prove the existence of black holes?}
\label{sec:5}

Horizon is the basic concept that defines a black hole: a massive object that consequently produces a gravitational attraction in the surrounding environment, but that has no material border. In fact, an invisible border in the space-time, which is predicted by general relativity, surrounds it. This way, matter could go through this border without being rejected, and without losing a fraction of its kinetic energy in a thermonuclear explosion, as sometimes is observed as x-ray bursts of type I  when the compact object is a neutron star instead of a black hole. In fact, as shown in Fig. \ref{fig:4}, the interval of time between the sudden drop of the flux and the spike in the X-ray light curve that marks the onset of the jet signaled by the starting rise of the infrared synchrotron emission  is of a few minutes, orders of magnitude larger than the dynamical time of the plasma in the inner accretion disk. Although the drop of the X-ray luminosity could be interpreted as dissipation of matter and energy, the most popular interpretation is that the hot gas that was producing the X-ray emission falls into the black hole, leaving the observable Universe. 

So, have we proved with such observations the existence of black holes? Indeed, we do not find any evidence of material borders around the compact object that creates gravitational attraction. However, the fact that we do not find any evidence for the existence of a material surface does not imply that it does not exist. In fact, such type I x-ray bursts are only observed in certain range of neutron star mass accretion rates. That means that it is not possible to prove the existence of black holes using the horizon definition. According to Saint Paul, \emph{``faith is the substance of hope for: the evidence of the not seen''}. That is why for some physicists, black holes are just objects of faith. Perhaps the intellectual attraction of these objects comes from the desire of discovering the limits of the Universe. In this context, studying the physical phenomena near the horizon of a black hole is a way of approaching the ultimate frontiers of the observable Universe. 

\section{The rotation\index{rotation} of black holes}
\label{sec:6}

For an external observer, black holes are the simplest objects in physics since they can be fully described by only three parameters: mass, rotation and charge. Although black holes could be born with net electrical charge, it is believed that because of interaction with environmental matter, astrophysical black holes rapidly become electrically neutral. The masses of black holes gravitating in binary systems can be estimated with Newtonian physics. However, the rotation is much more difficult to estimate despite it being probably the main driver in the production of relativistic jets. 

There is now the possibility of measuring the rotation of black holes by at least three different methods: a) X-ray continuum fitting \cite{Zhang,McClintock}, b) asymmetry of the broad component of the Fe K$_\alpha$ line from the inner accretion disk \cite{Tanaka}, and c)  quasi-periodic oscillations with a maximum fix frequency observed in the X-rays \cite{RemillardMcClintock}.  The main source of errors in the estimates of the angular momentum resides in the uncertainties of the methods employed.

The side of the accretion disk that is closer to the black hole is hotter and produces huge amounts of thermal X and $\gamma$ radiations and is also affected by the strange configuration of space-time. Indeed, next to the black hole, space-time is curved by the black hole mass and dragged by its rotation. This produces vibrations that modulate the X-ray emission. Studies of those X-ray continuum and vibrations suggest  that the microquasars that produce the most powerful jets are indeed those that are rotating fastest. It has been proposed that these pseudo-periodic oscillations in microquasars are, moreover, one of the best methods today to probe by means of observations general relativity theory in the limit of the strongest gravitational fields. 

Analogous oscillations in the infrared range, may have been observed in the super massive black hole at the centre of the Milky Way. The quasi-periods of the oscillations (a few milliseconds for the microquasars X-ray emission and a few tens of minutes for the galactic centre black hole infrared emission) are proportionally related to the masses of the objects, as expected from the physical analogy between quasars and microquasars. Comparing the phenomenology observed in microquasars to that in black holes of all mass scales, several correlations among observables such as among the radiated fluxes in the low hard X-ray state, quasi-periodic oscillations, flickering frequencies, etc., are being found and used to derive the mass and angular momentum, which are the fundamental parameters that describe astrophysical black holes. 


\section{Extragalactic microquasars, microblazars\index{microblazars} and ultraluminous X-ray sources}
\label{sec:7}

Have microquasars been observed beyond the Milky Way galaxy? X-ray satellites are detecting far away from the centers of external galaxies large numbers of compact sources called `ultraluminous X-ray sources', because their luminosities seem to be greater than the Eddington limit for a stellar-mass black hole \cite{Fabbiano}. Although a few of these sources could be black holes of intermediate masses of hundreds to thousands solar masses, it is believed that the large majority are stellar-mass black hole binaries. 

Since the discovery of quasars in 1963, it was known that some quasars could be extremely bright and produce high energetic emissions in a short time. These particular quasars are called blazars and it is thought that they are simply quasars whose jets point close to the Earth's direction. The Doppler effect produces thus an amplification of the signal and a shift into higher frequencies. With Rodr\'\i guez we imagined in 1999 the existence of microblazars, that is to say X-ray binaries where the emission is also in the Earth's direction \cite{MirabelRodriguez1999}. Microblazars may have been already observed but the fast variations caused by the contraction of the time scale in the relativistic jets, make their study very difficult. In fact, one question at the time of writing this chapter is whether microblazars could have been already detected as ``fast black hole X-ray novae'' \cite{Kasliwal}. In fact, the so called ``fast black hole X-ray novae'' Swift~J195509.6+261406\index{Swift~J195509.6+261406} (which is the possible source of GRB~070610 \cite{Kasliwal}), and V4541~Sgr \cite{Orosz} are compact binaries that appeared as high energy sources with fast and intense variations of flux, as expected in microblazars \cite{MirabelRodriguez1999}.

Although some fast variable ultraluminous X-ray sources could be microblazars, the vast majority do not exhibit the intense, fast variations of flux expected in relativistic beaming. Therefore, it has been proposed \cite{King} that the large majority are stellar black hole binaries where the X-ray radiation is --as the particle outflows-- anisotropic, but not necessarily relativistically boosted. In fact, the jets in the Galactic microquasar SS~433\index{SS~433}, which are directed close to the plane of the sky, have kinetic luminosities of more than a few times 10$^{39}$ erg/sec,  which would be super-Eddington for a black hole of 10 solar masses. 

An alternative model is that ultraluminous X-ray sources may be compact binaries with black holes of more than 30 solar masses that emit largely isotropically with no beaming into the line of sight, either geometrically or relativistically \cite{Pakull}. This conclusion is based on the formation, evolution and overall energetics of the ionized nebulae of several 100 pc diameter in which some ultraluminous X-ray sources are found embedded. The recent discoveries of high mass binaries with black holes of 15.7 solar masses in M~33\index{M~33} \cite{Orosz2007} and 23-34 solar masses in IC~10\index{IC~10} \cite{Prestwich} support this idea. Apparently, black holes of several tens of solar masses could be formed in starburst galaxies of relative low  metal content.

\section{Very energetic $\gamma$-ray emission from compact binaries}
\label{sec:8}

\begin{figure}
\centering
\includegraphics[height=6cm]{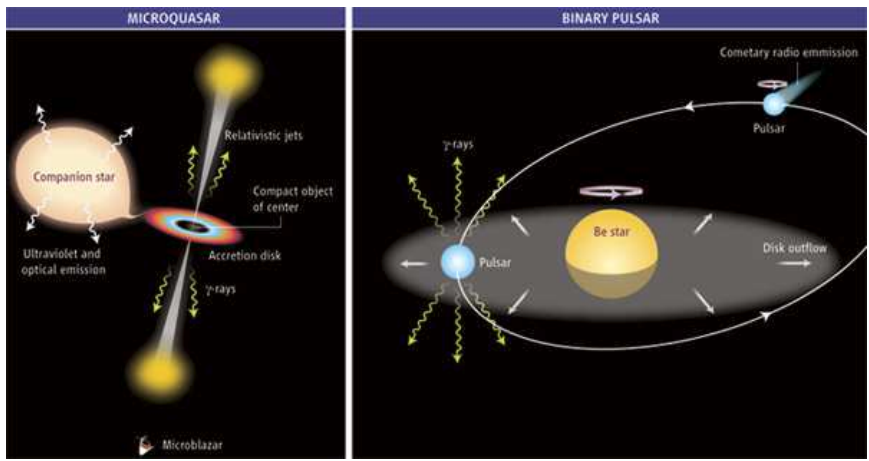}
\caption{
Alternative contexts for very energetic $\gamma$-ray binaries \cite{Mirabel2006}. Left: microquasars are powered by compact objects (neutron stars or stellar-mass black holes) via mass accretion from a companion star. The interaction of collimated jets with the massive outflow from the donor star can produce very energetic $\gamma$-rays by different alternative physical mechanisms \cite{Romero}, depending on whether the jets are baryonic or purely leptonic. Right: pulsar winds are powered by rotation of neutron stars; the wind flows away to large distances in a comet-shape tail. Interaction of this wind with the companion-star outflow may produce very energetic $\gamma$-rays \cite{Dubus}. 
}
\label{fig:5}
\end{figure}

Very energetic $\gamma$-rays with energies greater than 100 gigaelectron volts have recently been detected with ground based telescopes from four high mass compact binaries \cite{Mirabel2006}. These have been interpreted by models proposed in the contexts represented in Fig. \ref{fig:5}. In two of the four sources the $\gamma$ radiation seems to be correlated with the orbital phase of the binary, and therefore may be consistent with the idea that the very high energy radiation is produced by the interaction of  pulsar winds with the mass outflow from the massive companion star \cite{Dubus,DhawanMQW}. The detection of TeV emission from the black hole binary Cygnus~X-1\index{Cyg~X-1} \cite{Albert} and the TeV intraday variability in M~87\index{M~87} \cite{Aharonian} provided support to the jet models \cite{Romero}, which do not require relativistic Doppler boosting as in blazars and microblazars . It remains an open question whether the $\gamma$-ray binaries LS~5039\index{LS~5039} and LS~I~+61~303\index{LS~I~+61~303} could be microquasars where the $\gamma$ radiation is produced by the interaction of the outflow from the massive donor star  with jets \cite{Romero} or pulsar winds \cite{Dubus}.

\section{Microquasars and gamma-ray bursts}
\label{sec:9}

It is believed that gamma-ray bursts of long duration (t$>$1 sec) mark the birth of black holes by core collapse of massive stars. In this context, microquasars that contain black holes would be fossils of gamma-ray burst\index{gamma-ray burst} sources of long duration, and their study in the Milky Way and nearby galaxies can be used to gain observational inside into the physics of the much more distant sources of gamma-ray bursts. Questions of topical interest are: a) do all black hole progenitors explode as very energetic hypernovae of type Ib/c? ; b) what are the birth places and nature of the progenitors of stellar black holes?

The kinematics of microquasars provide clues to answer these questions. When a binary system of massive stars is still gravitationally linked after the explosion of one of its components, the mass centre of the system acquires an impulse, whatever matter ejection is, symmetric or asymmetric. Then according to the microquasar movement we can investigate the origin and the formation mechanism of the compact object. Knowing the distance, proper motion, and radial velocity of the centre of mass of the binary, the space velocity and past trajectory can be determined. Using multi-wavelength data obtained with a diversity of observational techniques, the kinematics of eight microquasars have so far been determined. 

\begin{figure}
\centering
\includegraphics[height=6cm]{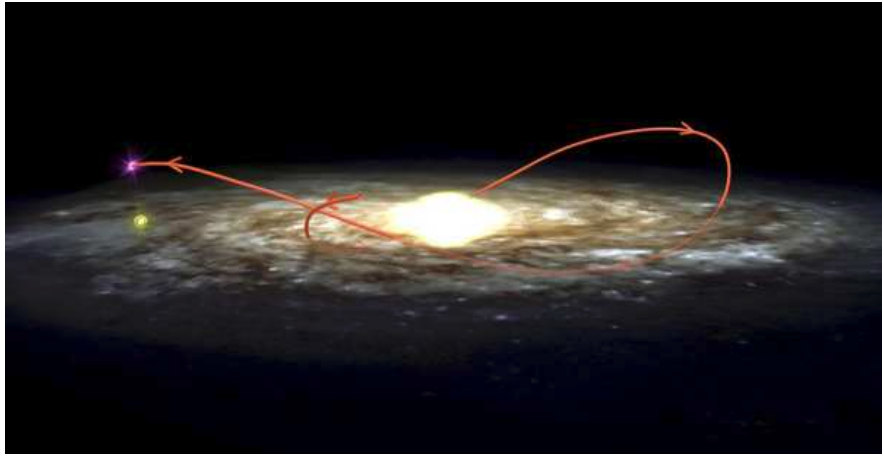}
\caption{
A wandering black hole in the Galactic halo \cite{Mirabel2001}. The trajectory of the black hole for the last 230 million years is represented in red. The bright dot on the left represents the Sun.  
}
\label{fig:6}
\end{figure}

One interesting case is the black hole wandering in the Galactic halo, which is moving at high speed, like globular clusters \cite{Mirabel2001} (Fig. \ref{fig:6}).  It remains an open question whether this particular halo black hole was kick out from the Galactic plane by a natal explosion, or is the fossil of a star that was formed more than 7 billions of years ago, before the spiral disk of stars, gas and dust of the Milky Way was formed. In this context, the study of these stellar fossils may represent the beginning of what could be called `Galactic Archaeology'. Like archaeologists, studying these stellar fossils, astrophysicists can infer what was the history of the Galactic halo.

The microquasars LS~5039\index{LS~5039} \cite{Ribo} and GRO~J1655-40\index{GRO~J1655-40} \cite{Mirabel2002} which contain compact objects with less that $\sim$7 solar masses were ejected from their birth place at high speeds, and  therefore the formation of these compact objects with relative small masses must have been associated with energetic supernovae. On the contrary, the binaries Cygnus~X-1\index{Cyg~X-1} \cite{Mirabel2003} and GRS~1915+105\index{GRS~1915+105} \cite{Dhawan2007} which contain black holes of at least 10 solar masses do not seem to have received a sudden impulse. Preliminary results on the kinematics of the X-ray binaries suggest that low mass black holes are formed by a delayed collapse of a neutron star with  energetic supernovae, whereas stellar black holes with masses equal or greater than 10 solar masses are the result of the direct collapse of massive stars, namely, they are formed in the dark. This is consistent with the recent finding of gamma-ray bursts of long duration in the near universe without associated luminous supernovae \cite{DellaValle}. 

There are indications that the mass of the resulting black hole may be a function of the metal content of the progenitor star. In fact, the black holes with 16 solar masses in M~33\index{M~33} \cite{Orosz2007} and more than 23 solar masses in IC~10\index{IC~10} \cite{Prestwich}, are in small galaxies of low metal content. This is consistent with the fact that the majority of the gamma-ray bursts of long duration take place in small starburst galaxies at high redshift, namely, in galactic hosts of low metal content \cite{LeFloch}. Since the power and redshift of gamma-ray bursts seem to be correlated this would imply a correlation between the mass of the collapsing stellar core and the power of the $\gamma$-ray jets.

\begin{figure}
\centering
\includegraphics[height=7.5cm]{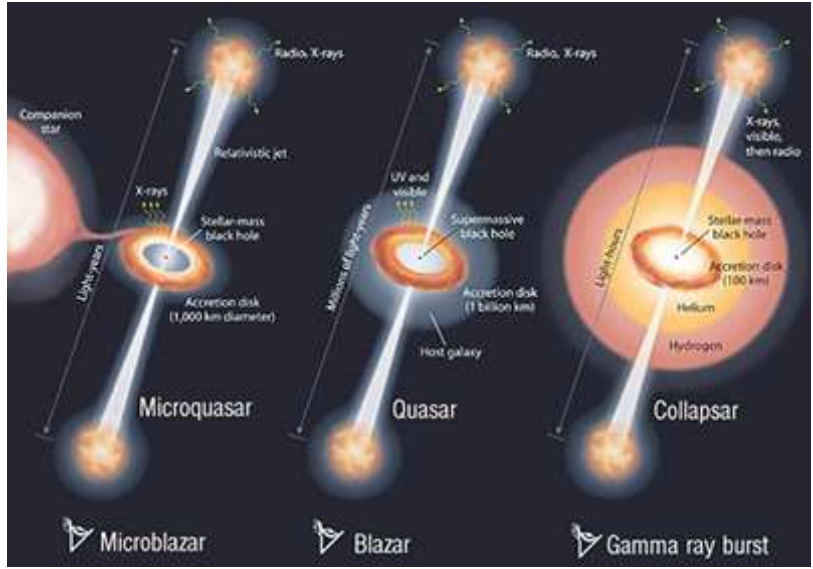}
\caption{
The same physical mechanism can be responsible for three different types of objects: microquasars, quasars and massive stars that collapse (`collapsars') to form a black hole producing gamma-ray bursts. Each one of these objects contains a black hole, an accretion disk and relativistic particles jets. Quasars and microquasars can eject matter several times, whereas the collapsars form jets only once. When the jets are aligned with the line of sight of the observer these objects appear as microblazars, blazars and gamma-ray bursts, respectively. Reproduced from \cite{MirabelSky}.  
}
\label{fig:7}
\end{figure}

Gamma-ray bursts of long duration are believed to be produced by ultra relativistic jets generated in a massive star nucleus when it catastrophically collapses to form a black hole. Gamma-ray bursts are highly collimated jets and it has been proposed \cite{MirabelSky} that there may be a unique universal mechanism to produce relativistic jets in the Universe, suggesting that the analogy between microquasars and quasars can be extended to the gamma-ray bursts sources, as illustrated in the diagram of Fig. \ref{fig:7}.

\section{Conclusions}
\label{sec:10}

Black-hole astrophysics is presently in an analogous situation as was stellar astrophysics in the first decades of the 20th century. At that time, well before the physical understanding of the interior of stars and the way by which they produce and radiate energy, empirical correlations such as the HR diagram were found and used to derive fundamental properties of the stars, such as the mass. Similarly, at present before a comprehensive understanding of black hole physics, empirical correlationsbetween X-ray and radio luminosities and characteristic time scales are being used to derive the    mass and spin of black holes of all mass scales, which are the fundamental parameters that describe astrophysical black holes. Therefore, there are historical and epistemological analogies between black hole astrophysics and stellar astrophysics.  
The research area on microquasars has become one of the most important areas in high energy astrophysics. In the last 14 years there have been seven international workshops on microquasars: 4 in Europe, 1 in America and 2 in Asia. They are currently attended by 100--200 young scientists who, with their work on microquasars, are contributing to open new horizons in the common ground of high energy physics and modern astronomy. 

{\bf Apologies:} This manuscript is based on short courses given at international schools  for graduate students, intended to give an introduction to this area of research. It is biased by my own personal choice, and hence it is by no means a comprehensive review.  Because references had to be minimized I apologize for incompleteness to colleagues working in the field. Part of this work was written while the author was staff member of the European Southern Observatory in Chile. 

%
%

%

\begin{thebibliography}{99.}
%
%
%

\bibitem{Aharonian} F.~Aharonian, A.G.~Akhperjanian, A.R., Bazer-Bachi et al.: Science, \textbf{314}, 1424 (2006)

\bibitem{Albert} J.~Albert, E.~Aliu, H.~Anderhub et al.: ApJ, \textbf{665}, L51 (2007)

\bibitem{Castro1915} A.J.~Castro-Tirado, S.~Brandt, N.~Lund et al.: ApJSuppl, \textbf{92}, 469 (1994)

\bibitem{DellaValle} M.~Della~Valle, G.~Chincarini, N.~Panagia et al.: Nature, \textbf{444}, 1050 (2006)

\bibitem{Dhawan} V.~Dhawan, I.F.~Mirabel, L.F. Rodr\'\i guez: ApJ, \textbf{543}, 373 (2000)

\bibitem{DhawanMQW} V.~Dhawan, A.~Mioduszewski, M.~Rupen: Proc. of the VI Microquasar Workshop: Microquasars and Beyond, PoS(MQW6)052 (2006)

\bibitem{Dhawan2007} V.~Dhawan, I.F.~Mirabel, M.~Rib\'o et al.: ApJ, \textbf{668}, 430 (2007)

\bibitem{Dubus} G.~Dubus: A\&A, \textbf{456}, 801 (2006)

\bibitem{Fabbiano} G.~Fabbiano: ARA\&A, \textbf{44}, 323 (2006)

\bibitem{Fender1999} R.P.~Fender, S.T.~Garrington, D.J.~McKay et al.: MNRAS, \textbf{304}, 865 (1999)

\bibitem{fbg} R.P.~Fender, T.M.~Belloni, E.~Gallo: MNRAS, \textbf{355}, 1105 (2004)

\bibitem{Giacconi} R.~Giacconi, H.~Rursky, J.R.~Waters: Nature, \textbf{204}, 981 (1964)

\bibitem{Greiner} J.~Greiner, J.G.~Cuby, M.J.~McCaughrean: Nature, \textbf{414}, 522 (2001)

\bibitem{Kasliwal} M.M.~Kasliwal, S.B.~Cenko, S.R.~Kulkarni et al.: ApJ, \textbf{678}, 1127 (2008)

\bibitem{King} A.R.~King, M.B.~Davies, M.J.~Ward et al.: ApJ, \textbf{552}, L109 (2001)

\bibitem{LeFloch} E.~Le Floc'h, P.-A.~Duc, I.F.~Mirabel et al.: A\&A, \textbf{400}, 499 (2003)

\bibitem{Leventhal} M.~Leventhal, C.J.~MacCallum, S.D.~Barthelmy et al.: Nature, \textbf{339}, 36 (1989)

\bibitem{Margon} B.~Margon, H.C.~Ford, J.I.~Katz et al.: ApJ, \textbf{230}, L41 (1979)

\bibitem{Marscher} A.P.~Marscher, S.G.~Jorstad, J.-L.~G\'omez: Nature, \textbf{417}, 625 (2002)

\bibitem{McClintock} J.E.~McClintock, R.~Shafee, R.~Narayan et al.: ApJ, \textbf{652}, 518 (2006)

\bibitem{Mirabel2006} I.F.~Mirabel: Science, \textbf{312}, 1759 (2006)

\bibitem{Mirabel1740} I.F.~Mirabel, L.F. Rodr\'\i guez, B.~Cordier et al.: Nature, \textbf{358}, 215 (1992)

\bibitem{MirabelRodriguez1994} I.F.~Mirabel, L.F. Rodr\'\i guez: Nature, \textbf{371}, 46 (1994)

\bibitem{Mirabel1998} I.F.~Mirabel, V.~Dhawan, S.~Chaty et al.: A\&A, \textbf{330}, L9 (1998)

\bibitem{MirabelRodriguez1998} I.F.~Mirabel, L.F. Rodr\'\i guez: Nature, \textbf{392}, 673 (1998)

\bibitem{MirabelRodriguez1999} I.F.~Mirabel, L.F. Rodr\'\i guez: ARA\&A, \textbf{37}, 409 (1999)

\bibitem{Mirabel2001} I.F.~Mirabel, V.~Dhawan, R.P.~Mignani et al.: Nature, \textbf{413}, 139 (2001)

\bibitem{MirabelSky} I.F.~Mirabel, L.F.~Rodr\'\i guez: Sky \& Telescope, \textbf{May 2002}, 32 (2002)

\bibitem{Mirabel2002} I.F.~Mirabel, R.P.~Mignani, I.~Rodrigues et al.: A\&A, \textbf{395}, 595 (2002)

\bibitem{Mirabel2003} I.F.~Mirabel, I.~Rodrigues: Science, \textbf{300}, 1119 (2003)

\bibitem{Orosz} J.A.~Orosz, E.~Kuulkers, M.~van der Klis et al.: ApJ, \textbf{555}, 489 (2001)

\bibitem{Orosz2007} J.A.~Orosz, J.E.~McClintock, R.~Narayan et al.: Nature, \textbf{449}, 872 (2007)

\bibitem{Pakull} M.W.~Pakull, L.~Mirioni: Rev. Mex. Astr. \& Ap., \textbf{15}, 197 (2003)

\bibitem{Prestwich} A.H.~Prestwich, R.~Kilgard, P.A.~Crowther et al.: ApJ, \textbf{669}, L21 (2007)

\bibitem{RemillardMcClintock} R.A.~Remillard, J.E.~McClintock: ARA\&A, \textbf{44}, 49 (2006)

\bibitem{Ribo} M.~Rib\'o, J.M.~Paredes, G.E.~Romero et al.: A\&A, \textbf{384}, 954 (2002)

\bibitem{Romero} G.E.~Romero: Gamma-Ray Emission from Microquasars: Leptonic vs. Hadronic Models. In: \textit{ Relativistic Astrophysics Legacy and Cosmology - Einstein's}, ESO Astrophysics Symposia (Springer-Verlag, Berlin Heidelberg 2008) pp 480--482

\bibitem{Tanaka} Y.~Tanaka: AN, \textbf{327}, 1098 (2006)

\bibitem{vdlaan} H.~van der Laan: Nature, \textbf{211}, 1131 (1966)

\bibitem{Weiden} G.~Weidenspointner, G.~Skinner, J.~Pierre et al.: Nature, \textbf{451}, 159 (2008)

\bibitem{Zhang} S.N.~Zhang, W.~Cui, W.~Chen: ApJ, \textbf{482}, L155 (1997)

\end{thebibliography}
%

\end{document}